\documentclass[11pt]{revtex4-2}
\usepackage{relsize}
\usepackage{braket}
\usepackage{hyperref}
\usepackage{amsmath,amsfonts,amssymb}
\hypersetup{dvips,dvipdfm,colorlinks=true,urlcolor=magenta,filecolor=magenta,linktoc=page,citecolor=red,linkcolor=blue,bookmarks=true}
\usepackage{graphicx,epstopdf}
\usepackage{bm}
\usepackage{hyperref}
\usepackage{amsmath,amsfonts,amssymb}

\hypersetup{dvips,dvipdfm,colorlinks=true,urlcolor=magenta,filecolor=magenta,linktoc=page,citecolor=red,linkcolor=blue,bookmarks=true}
\usepackage{graphicx,epstopdf}
\usepackage{array}

\newcounter{defcounter}
\newcolumntype{L}[1]{>{\raggedright\let\newline\\\arraybackslash\hspace{0pt}}m{#1}}
\newcolumntype{C}[1]{>{\centering\let\newline\\\arraybackslash\hspace{0pt}}m{#1}}
\newcolumntype{R}[1]{>{\raggedleft\let\newline\\\arraybackslash\hspace{0pt}}m{#1}}

\begin{document}
	
	\title{The canonical quantization of the cosmic fluid in the pre-inflationary era : a new kind of Emergent Universe.   \\
	}
	\author{Subhayan Maity\footnote {maitysubhayan@gmail.com}}
	\affiliation{Department of Mathematics, Jadavpur University, Kolkata-700032, West Bengal, India.}

	\author{Sujayita Bakra\footnote {sulukhejuri@gmail.com}}
	\affiliation{Department of Physics, Diamond Harbour Women University, West Bengal- 743388, India.}

	\begin{abstract}
		
		The early phase of cosmic evolution is still a great topic of interest to the cosmologist community. Especially the mechanism behind the origin of the Universe is yet to be resolved. Most people believe in the Big Bang models (singular origin of the Universe), but fewer of the cosmologists support in favor of  the non-singular origin of the Universe in their works. These models are based on the standard general theory of relativity. In this work, we have aimed to investigate the nature of the early cosmic evolution from the canonical quantization of the cosmic fluid and importantly to find a probable answer of what the true origin of the Universe is.

	\end{abstract}

	\maketitle
	
	
	
	
	\section{Introduction}
	People are still interested to explore what the real origin of the Universe is and how it evolves in the epoch immediately after its origin. Big Bang scenario is the most accepted model to describe the origin of the Universe till the recent cosmology. Here the main issue is that at the very beginning, the Universe was singular and hence no law of physics can be applicable at that point. Some  cosmological parameters diverge in the singular origin. 
	\par Nowadays, the emergent scenario of the cosmic origin is also becoming a matter of interest. Here the Universe is assumed to start from a non-zero constant volume \cite{Banerjee:2007qi,Bhattacharya:2016env,Bose:2020xml,Chakraborty:2014ora,Ellis:2002we,Ellis:2003qz,Guendelman:2014bva} from the time immemorial $(t\rightarrow-\infty)$.
	\par So far general theory of relativity has  no confirmed answer about the origin of the Universe. Hence it is worth exploring whether the emergent origin of the Universe is acceptable or not  from some other  aspects of the cosmic fluid. 
	\par  In some previous works\cite{Maity:2021mlx,Maity:2022gby}, S.Maity in collaboration with S.Chakraborty has already  successfully demonstrated the existence of emergent Universe under diffusion mechanism. The diffusion of cosmic fluid is a consequence of the non-equilibrium thermodynamic evolution of the Universe. 
	The interaction of the cosmic fluid with some cosmological scalar field \cite{Benisty:2017eqh,Benisty:2017lmt,Benisty:2017rbw,Benisty:2018oyy} leads to determine the cosmic evolution process in the framework of general relativity.
	There a particular choice of the time dependent scalar field has been proposed and the existence of an emergent Universe has been successfully exhibited. Diffusion is a macroscopic process and based on the phenomenological arguments. Microscopically the diffusion mechanism in the cosmological context is driven by the particle creation and annihilation mechanism of the cosmic fluid particles. Quantum field theory covers this particle creation- annihilation phenomena by the canonical quantization of the Hamiltonian of the cosmic fluid.
	\par This work is such an attempt to setup an cosmic fluid model with real Klein - Gordon field and to investigate the early phase of the cosmic evolution from the condition of the canonical quantization of the cosmic fluid. Also this model will aim to address the mechanism behind the cosmic particle creation and annihilation.

	\section{Quantization of real scalar field in FLRW space-time and its consequences }
	In the framework of Einstein's general relativity, the evolution of the cosmic fluid is governed by Friedmann equations in a flat FLRW Universe $ds^2= d t^2- a^2(t)(dx^2 +dy^2 +d z^2)$, $ a$ is the time dependent scale factor of the Universe.  Considering the cosmic fluid as a barotropic fluid with barotropic index $W$, the trivial from of Friedmann equations are
	\begin{equation}
		3 H^2= k \rho, ~ 2\dot{H}+3H^2 = -kP , \label{1}
	\end{equation}
	where $H=\frac{\dot{a}}{a}$, Hubble parameter of the Universe. $P$ and $\rho$ are the thermodynamic pressure and energy density of the cosmic fluid respectively. Effectively the conservation equation of the fluid will be as,
	\begin{equation}
		\dot{\rho}+3(1+W)H \rho =0  \label{2}.
	\end{equation} 
	In this work, the dynamics of the cosmic fluid is aimed to explore from the ideas of quantum field theory. Here we start from representing the cosmic fluid by the Lagrangian of real scalar field (Klein-Gordon field). It is the most trivial field with no charge symmetry. It holds neither a rigid nor a gauge symmetry.  In flat space-time (Minkwoski geometry), the form of the Lagrangian density is given by,
	\begin{equation}
		\mathcal{L}= \frac{1}{2} \partial _{\mu} \phi \partial ^{\mu}\phi-\frac{1}{2} m^2 \phi ^2, \label{3}
	\end{equation} with $m$, the mass of the particle. $\phi^* =\phi$ is a real scalar field of space-time. This Lagrangian density is Lorentz invariant. This $\mathcal{L}$ yields the equation of motion (standard Klein-Gordon equation) as
\begin{equation}
	(\Box ^2 +m^2) \phi =0,  \label{4}
\end{equation}
where $\Box^2=\partial_{\mu}\partial^{\mu}$, the D' Alembertier operator. The standard KG equation follows Lorentz covariance. 
\par 	Under the curved space-time geometry, it can be modified in the form
\begin{equation}
	\mathcal{L}= \frac{1}{2} \nabla_{\mu}\phi\nabla^{\mu}\phi -\frac{1}{2} (m^2+\zeta R) \phi ^2 , \label{5}
\end{equation}
$\nabla_{\mu}$ represents the covariant derivative. $R$ is the Ricci scalar and $\zeta$ is the coupling parameter. For minimal coupling, we have chosen $\zeta =0$ . This modified Lagrangian density is also a Lorentz invariant quantity. The Euler- Lagrangian equation yields the K-G equation from this Lagrangian density of equation (\ref{5}),
	\begin{equation}
		\left(\nabla_{\mu}\nabla^{\mu} + m^2 \right)\phi(X). =0 \label{6}
	\end{equation}
	$X=x^{\mu}=(t,\vec{x})$ is the space-time four vector.
	For flat FLRW metric, equation (\ref{6}) takes the form
\begin{equation}
	\ddot{\phi}+3H \dot{\phi} - \nabla^{\prime  ^2} \phi+ m^2 \phi =0 .\label{7}
\end{equation}

$H=\frac{\dot{a}}{a}$, the Hubble parameter.  $\vec{\nabla}^{\prime}=\frac{1}{a}\vec{\nabla}$. Let consider the formal solution of the equation (\ref{3}) in terms of the Fourier transform $\phi (X)=\int d^3k \tilde{\phi}(K)e^{-i\int \tilde{K} dX}$ . $\tilde{K}=\tilde{k}^{\mu}=(k^0,\vec{k})$, the effective four momenta in this case.
This modified KG equation does not hold Lorentz covariance. So it exhibits spontaneous Lorentz symmetry breaking.
The term $3H\dot{\phi}$ causes the dissipation of energy from  the real scalar field like a damped oscillator. 
	\par Here we adopt the modification of the Lagrangian of the cosmic fluid field only due to the altered metric as a manifestation of the gravity. Here we haven't  impose any extra term containing the signature of coupling between the real scalar field and the gravity. This approach is justified because we aimed to focus on the dynamics of the cosmic fluid under curved space-time.  

 Let consider the formal solution of the equation (\ref{3}) in terms of the Fourier transform $\phi (X)=\int d^3k \tilde{\phi}(K)e^{-i\int \tilde{K} dX}$ . $\tilde{K}=\tilde{k}^{\mu}=(k^0,\vec{k})$, the effective four momenta in this case.
The solutions of this K-G equation are found as,

\begin{equation}
	\phi(X)=\frac{1}{(2 \pi)^{\frac{3}{2}}} \int\frac{1}{\sqrt{2 \omega_0}} d^3k^{\prime}\left[ \mathcal{A}(\vec{k}^{\prime},t) e^{-i\int KdX} +\mathcal{A}^* (\vec{k}^{\prime},t)e^{+i\int KdX}  \right], \label{8}
\end{equation} 
where $ \mathcal{A}(\vec{k^{\prime}},t)=e^{-\frac{3}{2}\int Hdt} \alpha (\vec{k^{\prime }})$ and  $ \mathcal{A}^*(\vec{k^{\prime}},t)=e^{-\frac{3}{2}\int Hdt} \alpha^* (\vec{k^{\prime }})$. Also $k^0=i\frac{3}{2}H \pm \omega (t)$, $\int KdX= \int \omega (t) dt - \vec{k}.\vec{x}$. \par Here we assume the slow variation of $k^0$ i.e. $
\frac{\dot{k}^0}{k^0}<<1$.  \par   One may introduce $\vec{x}\rightarrow \vec{x}^{\prime}=a \vec{x}, \vec{k}\rightarrow \vec{k}^{\prime}=\frac{\vec{k}}{a}$. $\vec{x}^{\prime}, \vec{k}^{\prime}$ are the comoving space-coordinate and momenta respectively.Eventually $\vec{k}.\vec{x}= \vec{k}^{\prime}.\vec{x}^{\prime}$.  Here $\omega(t)= \sqrt{|\omega_0 ^2 - \frac{9}{4} H^2|},  ~\omega_0=\sqrt{k^{\prime 2}+m^2}$ .

\par Generally the comoving momenta $k^{\prime}$ is dependent on time. Hence $\omega_0$ is also time dependent.

\par Evidently, The solution of K-G equation is the  superposition of infinite numbers of damped harmonic oscillators with momentum values ranging $-\infty <k^{\prime}< \infty$.

	\par For quantization of the Hamiltonian, this solution of the field will be replaced by the field operator,
	\begin{equation}
		\hat{\phi}(X)=\frac{1}{(2 \pi)^{\frac{3}{2}}} \int\frac{1}{\sqrt{2 \omega_0}} d^3k^{\prime}\left[ \hat{A}(\vec{k},t) e^{-i\int KdX} +\hat{A^{\dagger}} (\vec{k},t)e^{+i\int KdX}  \right].     \label{9}
	\end{equation}
	The Hamiltonian in this case will be in the  trivial form (similar  as in  the Minkwoski space-time),
	\begin{equation}
			\hat{h}=\frac{1}{2}\int d^3 x^{\prime} \left[{\dot{
				{
					\hat{\phi}}}}^2+|\vec{\nabla}^{\prime}\hat{\phi}|^2\ + m^2 \hat{\phi} ^2 \right]  \label{10}
	\end{equation}
	The form of the Hamiltonian as in the  equation (\ref{10}) looks very trivial in this form but the explicit form of $\dot{\hat{\phi}}$  in FLRW metric adds some non -trivial terms in the Hamiltonian. 
	\par The form of the normal ordered Hamiltonian in this scenario can be found as,
	\begin{tiny}
		
		\begin{equation}
			:\hat{h}:=\int  \omega_0 ~ d^3 k\{ \hat{A}^{\dagger}(\vec{k},t)\hat{A}(\vec{k},t)\} +\left( \frac{9 H^2}{2} + i3H \omega \right ) \int  \frac{d^3k}{2 \omega_0} \hat{A}(\vec{k},t) \hat{A}(-\vec{k},t)  \ e^{-2i\int \omega dt}+  \left( \frac{9 H^2}{2} - i3H \omega \right ) \int  \frac{d^3k}{2 \omega_0} \hat{A}^{\dagger}(\vec{k},t) \hat{A}^{\dagger}(-\vec{k},t)  \ e^{+2i\int \omega dt}       \label{11} 
		\end{equation}
		
	\end{tiny}
	
	i.e.
	$:\hat{h}:=\int  \omega_0 ~ d^3 k\{ \hat{A}^{\dagger}(\vec{k},t)\hat{A}(\vec{k},t)\} +\sqrt{9H^2 \omega^2 +\frac{81}{4}H^4} ~e^{+ i \theta} \int  \frac{d^3k}{2 \omega_0} \hat{A}(\vec{k},t) \hat{A}(-\vec{k},t)  \ e^{-2i\int \omega dt}+  \sqrt{9H^2 \omega^2 +\frac{81}{4}H^4}~e^{- i \theta} \int  \frac{d^3k}{2 \omega_0} \hat{A}^{\dagger}(\vec{k},t) \hat{A}^{\dagger}(-\vec{k},t)  \ e^{+2i\int \omega dt}  $, with $\theta= \tan^{-1}(\frac{2w}{3H})$.
	\par 
	The first term in the right hand side $ \hat{h}_0 =\int   \omega_0 d^3 k\{ \hat{A}^{\dagger}(\vec{k},t)\hat{A}(\vec{k},t)\}$ of the equation (\ref{11}) is perfectly quantized with $\hat{A}^{\dagger}(\vec{k},t)$ and $\hat{A} (\vec{k},t)$ are the creation and annihilation operator respectively of the boson particle with comoving momenta $\vec{k}$.
	
	\begin{equation}
		\hat{A}^{\dagger}(\vec{k},t) \ket{0}= \ket{\vec{k}},\hat{A}(\vec{k},t) \ket{0}=0,  \label{12}
	\end{equation} 
	
	where $\ket{0}$ is vacuum state or zero particle state.  The corresponding number operator can be defined as
	\begin{equation}
		\hat{N}(\vec{k},t)=\hat{A}^{\dagger}(\vec{k},t)\hat{A}(\vec{k},t)    \label{13}
	\end{equation} 
	
	The second term in the expression of the Hamiltonian in the equation(\ref{11}) can not be quantized.

	In this case, we have several alternative restrictions on the parameters to  handle the non-trivial term and get the quantized form of the Hamiltonian operator.
	\paragraph{Case : $1$} $(H=0)$ 
	
	The free field canonical  quantization of the Hamiltonian is possible at an epoch of time when $H=0, a\neq 0$. 
	
	\par Under this condition, the form of the normal ordered Hamiltonian will be 
	\begin{equation*}
		\boxed{\hat{:h:}=\frac{1}{a^3} :\hat{h}_{\mbox{Minkwoski}} :,}
	\end{equation*}
		where $:\hat{h}_{\mbox{Minkwoski}} :$ is the normal ordered Hamiltonian  of the real K-G field in Minkwoski space-time.

	\paragraph{Case : $2$} $(H\neq 0)$   
	\par  In this case we have two options for quantization of the Hamiltonian.
	\par $(i)$ The free field quantization is also possible at any epoch of time (not necessarily for $H=0$) if
\begin{equation*}
	\boxed{\mbox{ Re} (\omega)=\frac{3}{2}H} .
\end{equation*}

\par $(ii)$	 Finally, if  $\mbox{Re}(\omega) \neq \frac{3}{2} H $ and $H$ is non-zero, then one may develop the quantization method analogs with the interacting field theory.  
Here the explicit form of $\dot{\phi}$ can be written as
\begin{equation}
	\dot{\hat{\phi}}(X)=e^{-\frac{3}{2}\int H dt} \left[\dot{\hat{\phi}}_0 - \frac{3}{2} H \hat{\phi} _0\right]=e^{-\frac{3}{2}\int H dt}\dot{\hat{\phi}}_0 - \frac{3}{2} H \hat{\phi}. \label{14}
\end{equation} 
$\hat{\phi}=e^{\pm\frac{3}{2}\int H dt} \hat{\phi}_0$  .

\par 
Consequently the form of Hamiltonian can be found as,
\begin{equation}
	\hat{h}=\int d^3 x \frac{1}{2} \left[ \dot{\hat{\phi}}_0 ^2 +  |\vec{\nabla}^{\prime}\hat{\phi}_0|^2 + m^2  \hat{\phi}_0^2            \right] + i\frac{3}{2}\omega H\int d^3x \left[\dot{\hat{\phi}}_0 \hat{\phi}_0 + \hat{\phi}_0 \dot{\hat{\phi}}_0   \right]  \label{15}
\end{equation}
Evidently, the Hamiltonian contains two parts, 
$\hat{h} = \hat{\mathcal{H}}_0 + \hat{\mathcal{H}}^{\prime} $. The first part $\hat{\mathcal{H}}_0 =\int d^3 x \frac{1}{2} \left[ \dot{\hat{\phi}}_0 ^2 +  |\vec{\nabla}^{\prime}\hat{\phi}_0|^2 + m^2  \hat{\phi}_0^2            \right]$ is the free field Hamiltonian and it can be quantized  with energy eigen value $\frac{\omega ^2}{\omega_0 a^3}$. But the second part $\hat{\mathcal{H}}^{\prime}$ can not be quantized and here we shall treat it as an interacting Hamiltonian. Hence we can calculate the $S$- matrix for this interacting Hamiltonian. 
\begin{equation}
	\hat{S}= T e^{-i \int_{\mathcal{T}}^{t} \hat{\mathcal{H}}^{\prime}_I(t)dt},\label{16}
\end{equation}  $T$ represents the time order product and $\hat{\mathcal{H}}^{\prime}_I(t)$  is the interacting Hamiltonian in interaction picture representation. 
\begin{equation}
	\hat{\mathcal{H}}^{\prime}_I(t)=  (i\frac{3}{2}\omega H)\int d^3x  \left[\dot{\hat{\phi}}_{0I} \hat{\phi}_{0I} + \hat{\phi}_{0I} \dot{\hat{\phi}}_{0I}   \right]. \label{17}
\end{equation}
Hence the spectrum at any epoch $t$ can be found from the relation 
\begin{equation}
	\ket{\vec{k},H(t)}_I= \hat{S} \ket{\vec{k},H(\mathcal{T})}_I, \label{18}
\end{equation}

	\section{ emergent scenario and cosmic evolution   }
	 
	 In this section, we shall determine the cosmic evolution pattern from the restrictions for the quantization of the cosmic fluid and highlight on the possibility of the existence of a non-singular origin of the Universe (emergent scenario).
	 \par From case$.1$, the conditions for free field quantization demand a particular time epoch where $H=0, a \neq 0$. This condition is identical with an emergent model of the Universe.  In this scenario, at the origin of the Universe (at infinite past), one has
	 
	 \cite{Banerjee:2007qi,Bhattacharya:2016env,Bose:2020xml,Chakraborty:2014ora,Ellis:2002we,Ellis:2003qz,Guendelman:2014bva,Maity:2022noc,Maity:2022lbq,Mukherjee:2005zt,Mukherjee:2006ds,Beesham:2009zw,Paul:2020bje,Zhang:2013ykz,Paul:2015eja,Debnath:2017xcu,Paul:2018ppy,Debnath:2020bno,Debnath:2021ncz,Paul:2022dsb,Paul:2021lvb,Paul:2010jb,Paul:2011nw,Ghose:2011fk,Labrana:2013oca,Paul:2019oxo}. 
	 \begin{equation}
	 	H\simeq 0, a\simeq a_E ~\mbox{when}  ~ t \rightarrow -\infty ,\label{19}
	 \end{equation} 
	 with $a_E$ is the scale factor of the Universe at the emergent epoch.

	 	Imposing this condition,  one has
	 \begin{equation}
	 	\hat{A}^{\dagger}(\vec{k},t) \rightarrow \left[a_E\right]^{- \frac{3}{2}} \alpha ^{\dagger}(\vec{k}), \hat{A}(\vec{k},t) \rightarrow \left[a_E\right]^{-\frac{3}{2}} \alpha (\vec{k}) . \label{20}
	 \end{equation}
	 Also the non-trivial terms vanishes at this epoch.
	 Hence one may find the Hamiltonian in the form of a free field Hamiltonian,
	 \begin{equation}
	 	\boxed{	:\hat{h}: = \hat{h}_0(a=a_E)=a_E^{-3} ~:\hat{h}_{\mbox{Minkwoski}} :} ,   \label{21}  
	 \end{equation} 
	 where $:\hat{h}_{\mbox{Minkwoski}} :$ is the normal ordered Hamiltonian  of the real K-G field in Minkwoski space-time. 
	 
	 Here the number operator will also be static with time,
	 \begin{equation}
	 	\boxed{	\hat{N}(\vec{k},t)= a_E^{- 3} \hat{N}(\vec{k})_{\mbox{Minkwoski}}.}    \label{22}
	 \end{equation}

	 \par  In case.$2$ $(ii)$ : ($H\neq 0, \mbox{Re}(\omega) \neq \frac{3}{2}H$), the Hamiltonian can be quantized in the method of interacting field theory. 
	 \par The free part of the Hamiltonian $\hat{\mathcal{H}_0}$ is in the canonically quantized form with energy eigen value $\epsilon _0 =\frac{\omega ^2}{\omega _0 a ^3}$. The interacting part of the Hamiltonian is given by 
	 
	 \begin{equation}
	 \mathcal{H}^{\prime}=	 i\frac{3}{2}\omega H\int d^3x \left[\dot{\hat{\phi}}_0 \hat{\phi}_0 + \hat{\phi}_0 \dot{\hat{\phi}}_0   \right]   \label{23}
	 \end{equation}

	 \par The system (the Universe) is free when $H=0,  \mathcal{H}^{\prime}=0 $ .
	 Again a valid free field Hamiltonian requires the conditions
	 $ H \simeq 0$ but $a\neq 0$.  Evidently these conditions are found at the infinite past (the origin of the emergent Universe),
	 $t \rightarrow-\infty, H\simeq 0, a \rightarrow a_E$. The energy eigen value of the free Hamiltonian leads to $\frac{\omega_0}{a_E^3}$. 
	 
	 \par Hence one may claim that both the free field quantization and interacting field quantization supports the emergent scenario of the Universe.
	 
	 At other epochs ($H\neq 0$), the quantization can be done in two ways. Firstly, the free field quantization is possible if $\mbox{Re}(\omega) = \frac{3}{2}H$. Hence only one specific energy eigen value and momentum value is acceptable $ \omega_0=\frac{3}{\sqrt{2}}H, 
	        k^{\prime}=\sqrt{\frac{9}{2}H^2-m^2}$. 
	 
	 In other case,  when  $\mbox{Re}(\omega) \neq \frac{3}{2}H$, the particle's state at any epoch can be related to the free field state (particle's state at $H\simeq 0$) through the $\hat{S}$ matrix. Let the free field state in interaction picture  is denoted as 
	 \begin{equation}
	 	\ket{free}_I=\lim_{t\rightarrow{-\infty}}\ket{\vec{k},H}_I .  \label{24}
	 \end{equation}
	 
	 Hence the state at any arbitrary epoch $t$ with Hubble parameter $H$ can be found as,
	 	\begin{equation}
	 	\ket{\vec{k},H}_I= \hat{S} \ket{free}_I, \label{37}
	 \end{equation}
	 
	 Here the form of $S$ matrix is as
	 	\begin{equation}
	 	\hat{S}= T e^{-i \int_{-\infty}^{t} \hat{\mathcal{H}}^{\prime}_I(t)dt}.\label{38}
	 \end{equation}

	 \begin{center}
		\begin{table}[!htb]
			\centering
			\renewcommand{\arraystretch}{2.5}
			\caption{} \label{tab:1}
			~~~~~~~\begin{tabular}{| >{\centering\arraybackslash}m{2.5cm}|>{\centering\arraybackslash}m{10cm}|>{\centering\arraybackslash}m{5cm}|}
				\hline
				Nature of quantization &  Epoch and Hubble parameter & Energy eigen value$(\epsilon)$ and restrictions \\
				\hline
				Free field Quantization & $t\rightarrow -\infty, a\rightarrow a_E, H\simeq 0$  & $\hat{h}=\frac{1}{a_E^3}\hat{h}_{\mbox{Minkwoski}}, \epsilon=\omega_0$, No restriction on $\omega_0$ \\

				\hline
				Free field quantization & $H\neq 0 $ & $ \mbox{Re}(\omega)=\frac{3}{2}H , \epsilon=\omega_0=\frac{3}{\sqrt{2}}H$\\
				\hline 
				Interacting field quantization & $t\rightarrow -\infty, a\rightarrow a_E, H\simeq 0$& $\epsilon=\frac{\omega_0}{a_E^3}$, No restriction on $\omega_0$  
				\\
				\hline
					Interacting field quantization & $H\neq 0$& $\omega \neq \frac{3}{2}H, \omega_0\neq \frac{3}{\sqrt{2}}H, ~~~\ket{
					\vec{k},H}_I=\hat{S}\ket{Free}_I$  
				\\
				\hline
				
			\end{tabular}
		\end{table}
	\end{center}
 \subsection{A complete pre-inflationary evolution model of the Universe.}
 At the origin ($t\rightarrow-\infty$) the canonical quantization of the cosmic fluid ensures the emergent (non-singular) origin of the Universe. But at the other epochs, there are two options to study the cosmic evolution dynamics from this quantization process. One may examine the free field quantization of the fluid particles with energy eigen value $\epsilon=\frac{3}{\sqrt{2}}H$ and momentum $k^{\prime}=\sqrt{\frac{9}{2}H^2-m^2}$. The particles with energy other than $\frac{3}{\sqrt{2}}H$ are quantized through an interaction at non-zero Hubble parameter epochs. But the outcomes of both free field and interacting field quantization yield the same cosmic evolution pattern.
 \par In this work we have chosen to examine the free field quantization process to obtain the evolution pattern of the Universe. Here we have phase - $1$ (say) with energy eigen value $\epsilon_1 = \omega_0$ at $H=0, a=a_E$  and phase - $2$ (say) with $\epsilon _2=\frac{3}{\sqrt{2}}H$ when $H\neq=0$. 
The phase -$2$ does not match with the phase -$1$ in the limit $H\rightarrow0$. 
\begin{equation*}
	\lim_{t\rightarrow -\infty}\epsilon_2\neq \epsilon_1.
\end{equation*}

 Hence we have proposed an intermediate phase (phase-$I$) to connect these two phases.
 \par The intermediate phase starts at the epoch with $H=\delta, \delta \rightarrow0$ and it lasts up to $H=\delta +\Delta $. $\Delta $ is also sufficiently small. Hence the energy eigen value of the cosmic fluid at this intermediate phase can be taken in the Taylor series about $H=\delta$ as,
 \begin{equation}
 	\epsilon_I(H)=\epsilon(\delta)+\left [\frac{\partial \epsilon_I}{\partial H} \right ]_{H=\delta} (H-\delta )+\frac{1}{2}\left [\frac{\partial^2 \epsilon_I}{\partial H^2} \right ]_{H=\delta} (H-\delta )^2 +...... \label{27}
 \end{equation}	
At the origin, the energy eigen value is supposed to be very high and it decreases immediately with expansion. Hence the slope of energy eigen value with Hubble parameter will be negative at the intermediate phase.

	 Considering the first order dependence on $H$ and $\delta\rightarrow0$, one may take 
	 \begin{equation}
	 	\epsilon_I(H)\simeq \epsilon(\delta)- \Gamma H, \label{28}
	 \end{equation}
 
 with $\left [\frac{\partial \epsilon_I}{\partial H} \right ]_{H=\delta}=-\Gamma$, a constant.  Now connecting with phase -$1$,
 \begin{equation*}
 	\lim_{t\rightarrow -\infty}\epsilon_I= \epsilon_1,
 \end{equation*}
we have $\epsilon(\delta)=\omega_0$ i.e. $\epsilon_I(H)=\omega_0-\Gamma H$.
The energy of each particle is decreasing with the expansion in the intermediate phase. Therefore one would expect the number of particle will be increasing. So in this particular phase, the cosmic fluid particles will be quantized as the growing Harmonic oscillators unlike the other phases of damped harmonic oscillators.

\par Again we have $\epsilon_I(\delta +\Delta )= \epsilon_2(\delta +\Delta)$. Hence we find $\delta +\Delta =\frac{\omega_0}{\Gamma +\frac{3}{\sqrt{2}}}$.
 This complete and continuous cosmic evolution before inflation has been described schematically in Fig-$1$.
\par In an isolated Universe, the total energy will be conserved. So one has the relation,
\begin{equation}
	<\hat{N}><\hat{h}>=E , \mbox{Constant.} \label{29}
\end{equation}
where $	<\hat{N}>=\bra{\epsilon(H),k^{\prime}(H)}\hat{N}\ket{\epsilon(H),k^{\prime}(H)}, <\hat{h}>=\bra{\epsilon(H),k^{\prime}(H)}\hat{h}\ket{H}=\epsilon(H)$. 
\par $\ket{\epsilon(H),k^{\prime}(H)}$ is the energy and momentum eigen state of the fluid particle with energy $\epsilon(H)$ and momentum $k^{\prime}(H)$.  $\hat{N}$ is the number operator $\hat{N}(k^{\prime},t)=\hat{A}^{\dagger}(k^{\prime},t)\hat{A}(k^{\prime},t)$. Let the eigenvalue of the number operator be $N(k^{\prime})$ i.e. 
\begin{equation*}
	\hat{N}(k^{\prime},t)\ket{\epsilon,k^{\prime}}=N(k^{\prime},t)\ket{\epsilon,k^{\prime}}.
\end{equation*}
Following the equation (\ref{29}), one gets
\begin{equation}
	\frac{\dot{N}}{N}+\frac{\dot{\epsilon}}{\epsilon}=0.   \label{30}
\end{equation}
\par  We can obtain the explicit form of the evolution equation in different phases from the equation (\ref{30}). The explicit forms of evolution equation and solutions are represented in the Table. $2$.

 \begin{center}
	\begin{table}[!htb]
		\centering
		\renewcommand{\arraystretch}{2.5}
		\caption{} \label{tab:1}
		~~~~~~~\begin{tabular}{| >{\centering\arraybackslash}m{3cm}|>{\centering\arraybackslash}m{1.5cm}|>{\centering\arraybackslash}m{6cm}|>{\centering\arraybackslash}m{7cm}|}
			\hline
			Phase of the evolution and ranges&  Energy eigen value$(\epsilon)$  & Number operator& Evolution equation and solution \\
			\hline
			PHASE-$1$,  $-\infty<t\leq t_1$ & $\epsilon_1=\omega_0$&Static oscillator :
			$\hat{A}(k^{\prime},t)=\left [a_E\right ]^{-\frac{3}{2}}\alpha(k^{\prime})$,$ \hat{A}^{\dagger}(k^{\prime},t)=\left [a_E\right ]^{-\frac{3}{2}}\alpha^{\dagger}(k^{\prime}), N(k^{\prime},t)=\left[a_E\right]^{-3} N_{\mbox{Minkowski}}$. & $a^{(1)}\simeq a_E, H^{(1)}\rightarrow0$ \\

			\hline
			PHASE -$I$, $t_1\leq t \leq t_2 $ & $\epsilon_I=\omega_0-\Gamma H $ &Growing oscillator : 	$\hat{A}(k^{\prime},t)=\left [a\right ]^{\frac{3}{2}}\alpha(k^{\prime}),~~~~~~ \hat{A}^{\dagger}(k^{\prime},t)=\left [a\right ]^{\frac{3}{2}}\alpha^{\dagger}(k^{\prime}), N(k^{\prime},t)=\left[a\right]^{3} N_{\mbox{Minkowski}}$.&$\dot{H}+3H\left(H-\frac{\omega_0}{\Gamma}\right )=0,
			a^{(I)}=a_2\left[\frac{\left (\frac{\alpha}{ H_2}-1 \right )+\exp 3 \alpha (t-t_2)}{\frac{\alpha }{ H_2}}\right]^{\frac{1}{3}},H^{(I)}=\alpha +\frac{H_2-\alpha }{a^3}, ~~\alpha =\frac{\omega_0}{\Gamma}.	$ \\
			\hline 
			PHASE-$2$, $t_2\leq t \leq t_0 $ &$\epsilon_2=\frac{3}{\sqrt{2}}H$&  Damped oscillator : $\hat{A}(k^{\prime},t)=\left [a\right ]^{-\frac{3}{2}}\alpha(k^{\prime}),~~~~~~ \hat{A}^{\dagger}(k^{\prime},t)=\left [a\right ]^{-\frac{3}{2}}\alpha^{\dagger}(k^{\prime}), N(k^{\prime},t)=\left[a\right]^{-3} N_{\mbox{Minkowski}}$.& $\dot{H}-3H^2=0, 	a^{(2)}=a_2\left[1-3H_0(t-t_2)\right]^{-\frac{1}{3}}, 	H^{(2)}=H_2\left(\frac{a}{a_2}\right)^3. $
			\\

			\hline
			
		\end{tabular}
	\end{table}
\end{center}

The continuity of the scale factor and the Hubble parameter (listed in Table $2$) at the transition epochs $t_1$ and $t_2$ ($t_1\rightarrow -\infty$) yields 
\begin{equation*}
	a_E=\left(1-\frac{\Gamma H_2}{\omega_0}\right)^{\frac{1}{3}}, \omega_0>\Gamma H_2 .
\end{equation*}

Here $H_1=H(t_1)=\delta, H_2=H(t_2)=\delta +\Delta $ and $a_2=a(t_2), a_1=a(t_1)$.

The continuous  variation of the scale factor and the Hubble parameter in different phases is shown in Fig. $2$.

\begin{figure}[h]
	\begin{minipage}{1\textwidth}
		\centering
		\includegraphics*[width=1\linewidth]{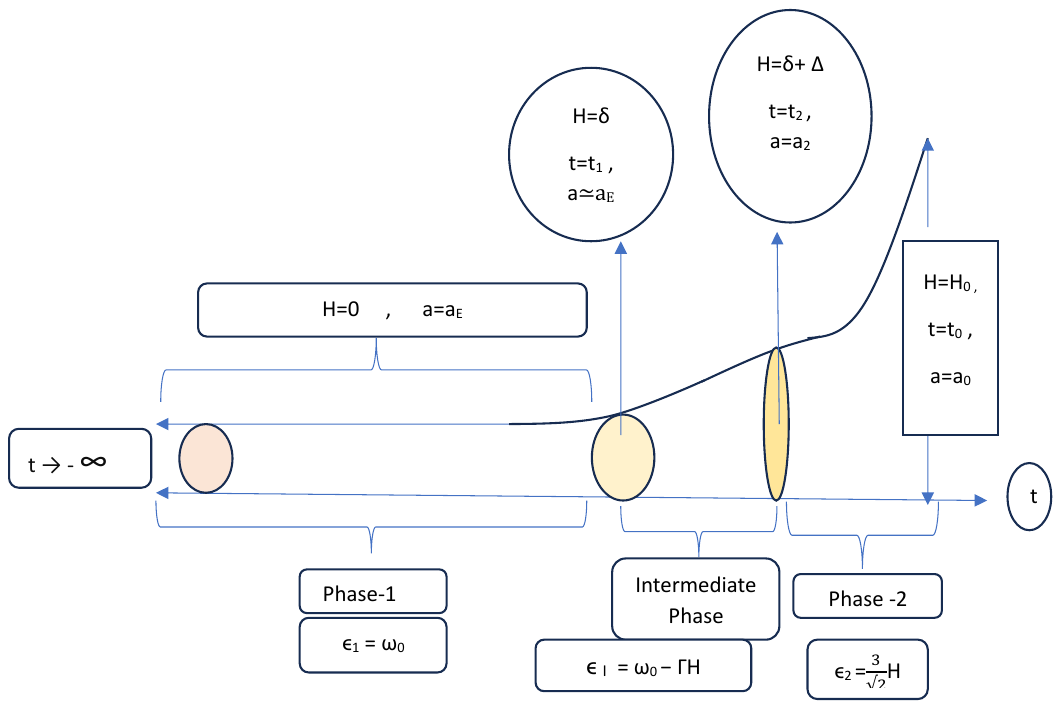}\\
		
	\end{minipage}
\begin{center}
	
		\caption{Pre-inflationary evolution in an emergent Universe under the canonical quantization of the cosmic fluid.} 
	\end{center}
	\label{fig1}
	
\end{figure}

	\begin{figure}[h]
	\begin{minipage}{0.49\textwidth}
		\centering
		\includegraphics*[width=0.9\linewidth]{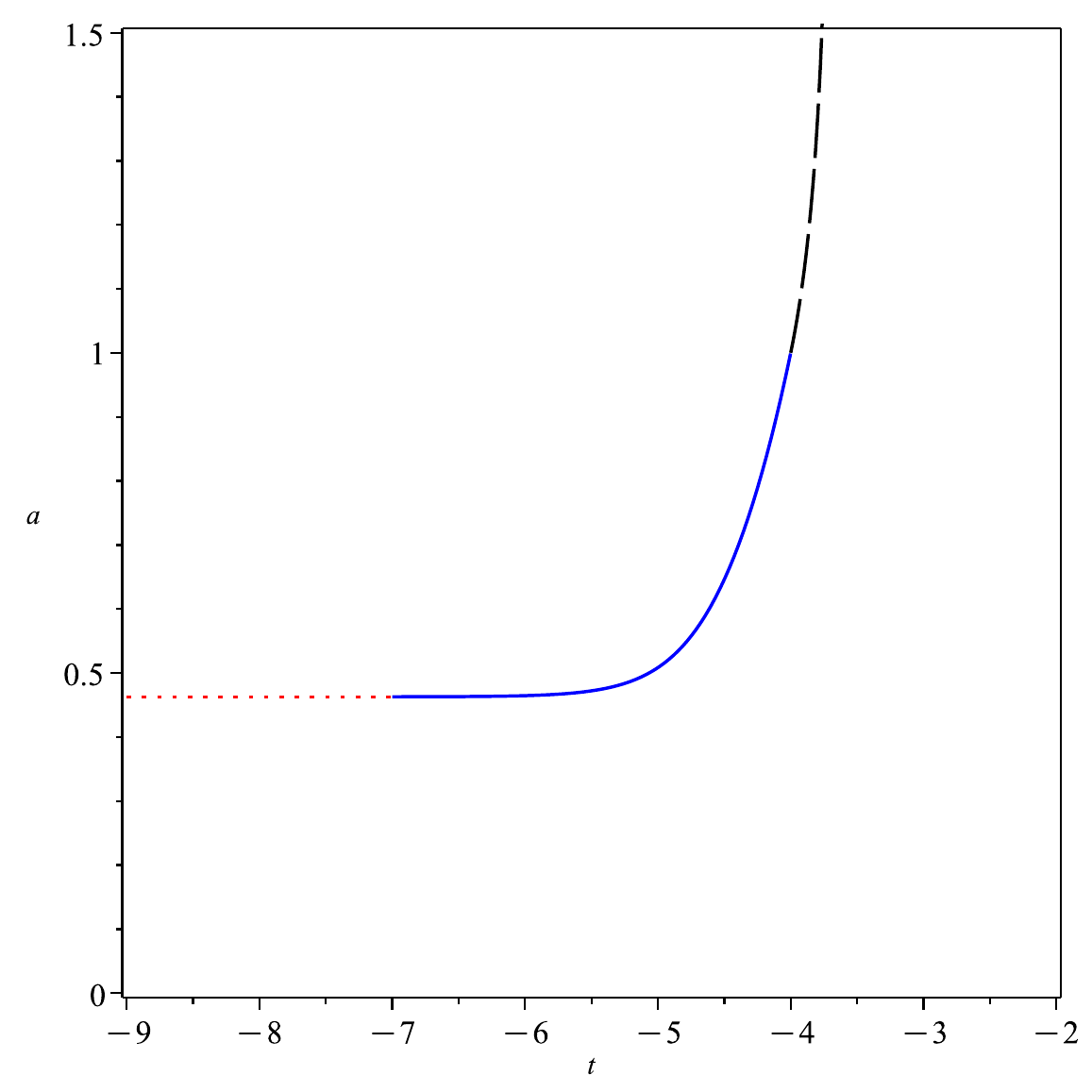}\\
		(a)
	\end{minipage}
	\begin{minipage}{0.49\textwidth}
		\centering
		\includegraphics*[width=0.9\linewidth]{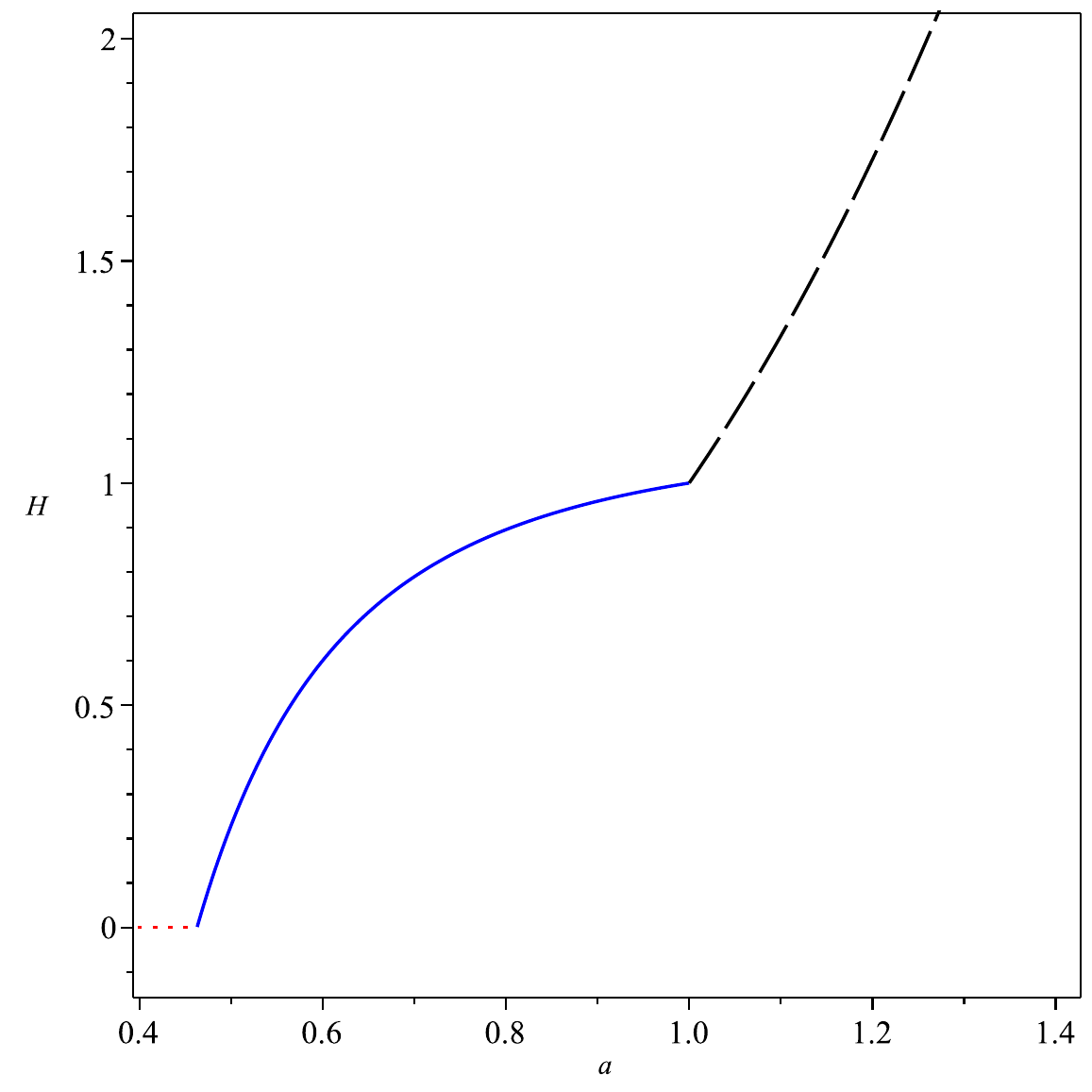}\\
		(b)
		
	\end{minipage}
	\begin{center}
		\caption{ (a)Variation of Scale factor  : $a$ with time $t$      (b) Variation of Hubble parameter : $H$with $a$,  for $t_1=-7,t_2=-4,H_2=1,a_2=1,\alpha =1.11$  (dot,red)- PHASE-$1$, (solid,blue)- PHASE - $I$, (long dash, black)-PHASE -$2$ .} 
	\end{center}
	\label{fig2}
	
\end{figure}

The phase -$2$ contains a big-rip singularity at the epoch $t_s=t_2+\frac{1}{3H}$. Here one may observe that at the limit of the time epoch $t_0 \rightarrow t_s^+$, $a$ starts to increase very rapidly. This may be considered as the  symptoms of transition to the inflationary era. So the Phase -$2$ terminates at some epoch before $t_s$ and the mathematical big-rip singularity has no relevance in the context of the cosmic evolution. Thus one may design the continuous cosmic evolution from an emergent era to the beginning of an inflationary era through these three phases from the canonical quantization process of the cosmic fluid.

\section{Discussion}

This work is a demonstration of the evolutionary model from the free field canonical quantization of the cosmic fluid. The quantization of the cosmic fluid Hamiltonian demands the non-singular origin of the Universe. Importantly in the pre-inflationary epoch, the quantization of the scalar field causes different evolutionary patterns in different phases. At the origin, the scalar field Hamiltonian remains quantized as the static (free) harmonic oscillators and yields static scale factor and zero Hubble parameter at the infinite past.   At other epoch ($H\neq 0$), the free field quantization is possible with only one specific energy and momentum value. Here the scalar field behaves as the damped oscillators. But these two phases are not continuous. In order to make connection between these two phases, we have introduced an intermediate phase. In this phase, the scalar field is quantized like the growing oscillators.   
\par Thus a smooth and continuous pre-inflationary evolution has been successfully established through these three phases.  Notably in a previous work, we have demonstrated an emergent model through a phenomenological choice of scale factor and Hubble parameter which are consistent with the quantization process of the cosmic fluid. This work is a different kind of emergent model and follows the direct causality of the free field quantization process.
\par  We especially want to highlight on the intermediate phase of this model. Such cosmic evolutionary phase is also found in other emergent models from diffusion mechanism, non-equilibrium thermodynamic mechanism etc.
Here we find that this phase is not the complete pre-inflationary phase, it is only the connection between the static phase and another pre-inflationary phase.  
\par We are hopeful that this finding will help to understand the physical mechanism of the  origin of the Universe. Also in future works, we aim to address the other aspects of cosmic evolution from the quantization process of the cosmic fluid.

	\section*{Acknowledgment } The author SM acknowledges Prof. Subenoy Chakraborty, Dept. of Mathematics, Jadavpur University, Kolkata-$700032$ for his valuable suggestion on  this topic and the author SB thanks all the faculties, Dept. of Physics, Diamond Harbour Women University for their assistance while this work.


\begin{thebibliography}{50}
		\bibitem{Chakraborty:2014ora}
		S.~Chakraborty,
		Phys. Lett. B \textbf{732}, 81-84 (2014)
		doi:10.1016/j.physletb.2014.03.028
		[arXiv:1403.5980 [gr-qc]].
		
		
		
		
		
		
		\bibitem{Ellis:2003qz}
		G.~F.~R.~Ellis, J.~Murugan and C.~G.~Tsagas,
		Class. Quant. Grav. \textbf{21}, no.1, 233-250 (2004)
		doi:10.1088/0264-9381/21/1/016
		[arXiv:gr-qc/0307112 [gr-qc]].
		
		\bibitem{Ellis:2002we}
		G.~F.~R.~Ellis and R.~Maartens,
		Class. Quant. Grav. \textbf{21}, 223-232 (2004)
		doi:10.1088/0264-9381/21/1/015
		[arXiv:gr-qc/0211082 [gr-qc]].
		
		
			\bibitem{Guendelman:2014bva}
			E.~Guendelman, R.~Herrera, P.~Labrana, E.~Nissimov and S.~Pacheva,
			Gen. Rel. Grav. \textbf{47}, no.2, 10 (2015)
			doi:10.1007/s10714-015-1852-1
			[arXiv:1408.5344 [gr-qc]].
		
		
		\bibitem{Bose:2020xml}
		A.~Bose and S.~Chakraborty,
		Phys. Dark Univ. \textbf{30}, 100740 (2020)
		doi:10.1016/j.dark.2020.100740
		[arXiv:2011.04649 [gr-qc]].
		
		\bibitem{Bhattacharya:2016env}
		S.~Bhattacharya and S.~Chakraborty,
		Class. Quant. Grav. \textbf{33}, no.3, 035013 (2016)
		doi:10.1088/0264-9381/33/3/035013
		[arXiv:1601.03816 [gr-qc]].
		
		
		\bibitem{Banerjee:2007qi}
		A.~Banerjee, T.~Bandyopadhyay and S.~Chakraborty,
		Grav. Cosmol. \textbf{13}, 290-292 (2007)
		[arXiv:0705.3933 [gr-qc]].
		\bibitem{Maity:2022gby}
		S.~Maity and S.~Chakraborty,
		Int. J. Mod. Phys. A \textbf{37}, no.03, 2250016 (2022)
		doi:10.1142/S0217751X22500166
		[arXiv:2302.07096 [gr-qc]].
		
		\bibitem{Maity:2021mlx}
		S.~Maity and S.~Chakraborty,
		Int. J. Mod. Phys. A \textbf{36}, no.29, 2150199 (2021)
		doi:10.1142/S0217751X21501992
		
		
		\bibitem{Benisty:2018oyy}
		D.~Benisty, E.~Guendelman and Z.~Haba,
		Phys. Rev. D \textbf{99}, no.12, 123521 (2019)
		[erratum: Phys. Rev. D \textbf{101}, no.4, 049901 (2020)]
		doi:10.1103/PhysRevD.99.123521
		[arXiv:1812.06151 [gr-qc]].
		
		\bibitem{Benisty:2017lmt}
		D.~Benisty and E.~I.~Guendelman,
		Int. J. Mod. Phys. A \textbf{33}, no.20, 1850119 (2018)
		doi:10.1142/S0217751X18501191
		[arXiv:1710.10588 [gr-qc]].
		
		
		\bibitem{Benisty:2017rbw}
		D.~Benisty and E.~I.~Guendelman,
		Int. J. Mod. Phys. D \textbf{26}, no.12, 1743021 (2017)
		doi:10.1142/S0218271817430210
		
		
		\bibitem{Benisty:2017eqh}
		D.~Benisty and E.~I.~Guendelman,
		Eur. Phys. J. C \textbf{77}, no.6, 396 (2017)
		doi:10.1140/epjc/s10052-017-4939-x
		[arXiv:1701.08667 [gr-qc]].
		
		
		
		
		
		
		\bibitem{Maity:2022noc}
		S.~Maity and S.~Chakraborty,
		Annals Phys. \textbf{444}, 169045 (2022)
		doi:10.1016/j.aop.2022.169045
		\bibitem{Maity:2022lbq}
		S.~Maity,
		[arXiv:2205.09759 [gr-qc]].
		
		
		
		\bibitem{Mukherjee:2005zt}
		S.~Mukherjee, B.~C.~Paul, S.~D.~Maharaj and A.~Beesham,
		[arXiv:gr-qc/0505103 [gr-qc]].
		
		
		
		\bibitem{Mukherjee:2006ds}
		S.~Mukherjee, B.~C.~Paul, N.~K.~Dadhich, S.~D.~Maharaj and A.~Beesham,
		Class. Quant. Grav. \textbf{23} (2006), 6927-6934
		doi:10.1088/0264-9381/23/23/020
		[arXiv:gr-qc/0605134 [gr-qc]].
		
		
		\bibitem{Beesham:2009zw}
		A.~Beesham, S.~V.~Chervon and S.~D.~Maharaj,
		Class. Quant. Grav. \textbf{26} (2009), 075017
		doi:10.1088/0264-9381/26/7/075017
		[arXiv:0904.0773 [gr-qc]].
		
		
		\bibitem{Paul:2020bje}
		B.~C.~Paul, S.~D.~Maharaj and A.~Beesham,
		[arXiv:2008.00169 [astro-ph.CO]].
		
		
		\bibitem{Zhang:2013ykz}
		K.~Zhang, P.~Wu and H.~Yu,
		JCAP \textbf{01} (2014), 048
		doi:10.1088/1475-7516/2014/01/048
		[arXiv:1311.4051 [gr-qc]].
		
		
		\bibitem{Paul:2015eja}
		B.~C.~Paul and A.~Majumdar,
		Class. Quant. Grav. \textbf{32} (2015) no.11, 115001
		doi:10.1088/0264-9381/32/11/115001
		[arXiv:1503.08284 [gr-qc]].
		
		
		\bibitem{Debnath:2017xcu}
		P.~S.~Debnath and B.~C.~Paul,
		Mod. Phys. Lett. A \textbf{32} (2017) no.39, 1750216
		doi:10.1142/S0217732317502169
		
		\bibitem{Paul:2018ppy}
		B.~C.~Paul and A.~S.~Majumdar,
		Class. Quant. Grav. \textbf{35} (2018) no.6, 065001
		doi:10.1088/1361-6382/aaa6a3
		
		
		\bibitem{Debnath:2020bno}
		P.~S.~Debnath and B.~C.~Paul,
		Int. J. Geom. Meth. Mod. Phys. \textbf{17} (2020) no.07, 2050102
		doi:10.1142/S0219887820501029
		
		\bibitem{Debnath:2021ncz}
		P.~S.~Debnath and B.~C.~Paul,
		Astrophys. Space Sci. \textbf{366} (2021) no.3, 32
		doi:10.1007/s10509-021-03937-3
		
		\bibitem{Paul:2022dsb}
		B.~C.~Paul, S.~D.~Maharaj and A.~Beesham,
		Int. J. Mod. Phys. D \textbf{31} (2022) no.06, 2250045
		doi:10.1142/S0218271822500456
		
		
		\bibitem{Paul:2021lvb}
		B.~C.~Paul,
		Eur. Phys. J. C \textbf{81} (2021) no.8, 776
		doi:10.1140/epjc/s10052-021-09562-2
		
		
		
		
		\bibitem{Paul:2010jb}
		B.~C.~Paul, P.~Thakur and S.~Ghose,
		Mon. Not. Roy. Astron. Soc. \textbf{407} (2010), 415
		doi:10.1111/j.1365-2966.2010.16909.x
		[arXiv:1004.4256 [astro-ph.CO]].
		
		\bibitem{Paul:2011nw}
		B.~C.~Paul, S.~Ghose and P.~Thakur,
		Mon. Not. Roy. Astron. Soc. \textbf{413} (2011), 686
		doi:10.1111/j.1365-2966.2010.18177.x
		[arXiv:1101.1360 [astro-ph.CO]].
		
		\bibitem{Ghose:2011fk}
		S.~Ghose, P.~Thakur and B.~C.~Paul,
		Mon. Not. Roy. Astron. Soc. \textbf{421} (2012), 20
		doi:10.1111/j.1365-2966.2011.19743.x
		[arXiv:1105.3303 [astro-ph.CO]].
		
		\bibitem{Labrana:2013oca}
		P.~Labra\~na,
		Phys. Rev. D \textbf{91} (2015) no.8, 083534
		doi:10.1103/PhysRevD.91.083534
		[arXiv:1312.6877 [astro-ph.CO]].
		
		\bibitem{Paul:2019oxo}
		B.~C.~Paul and A.~Chanda,
		Gen. Rel. Grav. \textbf{51} (2019) no.6, 71
		doi:10.1007/s10714-019-2551-0
		
		
		
		
		
		
		
		
		\bibitem{Perez:2020cwa}
		A.~Perez, D.~Sudarsky and E.~Wilson-Ewing,
		Gen. Rel. Grav. \textbf{53}, no.1, 7 (2021)
		doi:10.1007/s10714-020-02781-0
		[arXiv:2001.07536 [astro-ph.CO]].
		
		
		
		
		
		
		
		\bibitem{Maity:2019knj} 
		S.~Maity, P.~Bhandari and S.~Chakraborty,
		Eur.\ Phys.\ J.\ C {\bf 79}, no. 1, 82 (2019)
		doi:10.1140/epjc/s10052-019-6603-0
		[arXiv:1902.08037 [gr-qc]].
		
		
		
		
		
		
		\bibitem{Muller:2012kv} 
		S.~Muller {\it et al.},
		Astron.\ Astrophys.\  {\bf 551}, A109 (2013)
		doi:10.1051/0004-6361/201220613
		[arXiv:1212.5456 [astro-ph.CO]].
		
		
		
		\bibitem{MacDougall} 
		S. R. de Groot, Thermodynamics of Irreversible Processes, North-Holland Publishing Comp., amsterdam, 1951,
		The Journal of Physical Chemistry.$242$ S. F. $17.50.$
		doi 10.1021/j150492a019
		
		
		\bibitem{Haba:2017hch}
		Z.~Haba,
		``Models of a Diffusive DM/DE Interaction,''
		Acta Phys. Polon. Supp. \textbf{10}, 333-337 (2017)
		doi:10.5506/APhysPolBSupp.10.333
		
		\bibitem{Zeldovich:1971mw}
		Y.~B.~Zeldovich and A.~A.~Starobinsky,
		Sov. Phys. JETP \textbf{34}, 1159-1166 (1972)
		
		
		
		\bibitem{Senfit}
		Senft, J. (2007). Mechanical Efficiency of Heat Engines. Cambridge: Cambridge University Press. doi:10.1017/CBO9780511546105
		
		
		
		\bibitem{Chakraborty:2017efi}
		S.~Chakraborty and S.~Bhattacharya,
		Int. J. Mod. Phys. D \textbf{27}, no.14, 1847019 (2018)
		doi:10.1142/S0218271818470193
		[arXiv:1711.08868 [gr-qc]].
		
		
		
		
	\end{thebibliography}
\end{document}